\documentstyle[12pt, epsfig]{article}
\newcommand{\bea}   {\begin{eqnarray}}
\newcommand{\eea}   {\end{eqnarray}}
\begin{document}
\renewcommand{\thefootnote}{\fnsymbol{footnote}}

\thispagestyle{empty}

\title{Decomposition and Oxidation of the $N$-Extended Supersymmetric Quantum Mechanics Multiplets}
\author{Zhanna Kuznetsova\thanks{{\em e-mail: zhanna@cbpf.br}}
~and Francesco Toppan\thanks{{\em e-mail: toppan@cbpf.br}}
\\ \\
{\it $~^\ast$ICE-UFJF, cep 36036-330, Juiz de Fora (MG), Brazil}
\\
{\it $~^\dagger$ CBPF, Rua Dr.}
{\it Xavier Sigaud 150,}
 \\ {\it cep 22290-180, Rio de Janeiro (RJ), Brazil}}
\maketitle
\begin{abstract}
We furnish an algebraic understanding of the inequivalent connectivities (computed up to $N\leq 10$)
of the graphs associated to the irreducible supermultiplets of the $N$-extended Supersymmetric Quantum Mechanics.
We prove that the inequivalent connectivities of the
$N=5$ and $N=9$ irreducible supermultiplets
are due to inequivalent decompositions into two sets of $N=4$ (respectively, $N=8$)
supermultiplets. ``Oxido-reduction" diagrams linking the irreducible supermultiplets of
the $N=5,6,7,8$ supersymmetries are presented. We briefly discuss these results and their possible applications.
\end{abstract}
\vfill
\rightline{CBPF-NF-010/07}

\newpage
\section{Introduction}

The superalgebra of the Supersymmetric Quantum Mechanics ($1D$ $N$-Extended Supersymmetry
Algebra) is given by $N$ odd generators $Q_i$ ($i=1,\ldots , N$) and a single even generator
$H$ (the hamiltonian). It is defined by the (anti)-commutation relations
\begin{eqnarray}\label{nsusyalgebra}
\{Q_i,Q_j\}&=& 2\delta_{ij} H,\nonumber\\
\relax [Q_i, H] &=& 0.
\end{eqnarray}
The knowledge of its representation theory is essential
for several applications ranging from understanding features of higher dimensional supersymmetric theories from their dimensional reduction to one-time dimension, to the construction
of off-shell invariant $1D$ supersymmetric sigma-models associated to some $d$-dimensional target manifold.\par
The structure of the irreducible linear representations of the (\ref{nsusyalgebra}) superalgebra has been substantially elucidated in recent years. On the other hand, some open questions remain. This paper addresses and solves one of these issues.\par
In this Introduction we briefly recall the state of the art concerning the irreducible linear representations of (\ref{nsusyalgebra}) and explain this paper's results. We
postpone to the Conclusions all discussions on their relevance and their possible applications.\par
We recall that the irreducible linear representations we are interested in (the irreducible supermultiplets) are given by a finite number of fields,
bosonic and fermionic, depending on a single coordinate $t$ (the time). The generator $H$ is represented by the time-derivative, while
the $Q_i$'s generators are linear operators (matrices) whose entries are either $c$-numbers or time-derivatives up
to a certain power.\par
The program of classifying such irreducible supermultiplets starts with \cite{pt}, whose main results
can be stated as follows. All irreducible representations of (\ref{nsusyalgebra}),
for a given $N$, are obtained by applying a dressing transformation to a fundamental irreducible representation (nowadays called in the literature the ``root multiplet"), with equal
number of bosonic and fermionic fields. The root multiplet is specified by an associated Clifford algebra.
As a  corollary, the total number $n$ of bosonic fields entering an irreducible representation (which equals the total number of fermionic fields) is expressed, for any given $N$,
 by the following relation \cite{pt}
\begin{eqnarray}\label{irrepdim}
N&=& 8l+m,\nonumber\\
n&=& 2^{4l}G(m),
\end{eqnarray}
where $l=0,1,2,\ldots$ and $m=1,2,3,4,5,6,7,8$.\par
$G(m)$ appearing in (\ref{irrepdim}) is the Radon-Hurwitz function
\begin{eqnarray}&\label{radonhur}
\begin{tabular}{|c|cccccccc|}\hline
$m $&$1$&$2$&$3$& $4$&$5$&$6$&$7$&$8$\\ \hline
$G(m)  $&$1$&$2$&$4$& $4$&$8$&$8$&$8$&$8$\\ \hline
\end{tabular}&\nonumber\\
&&\end{eqnarray}
\par
A mass-dimension $d$ can be assigned to any field entering a linear representation
(the hamiltonian $H$ has a mass-dimension
$1$). Bosonic (fermionic) fields have integer (respectively, half-integer) mass-dimension. Each finite
linear representation is characterized by its ``fields content", i.e. the set of integers $(n_1,n_2,\ldots , n_l)$
specifying the number $n_i$ of fields of dimension
$d_i$ ($d_i = d_1 + \frac{i-1}{2}$, with $d_1$ an arbitrary constant) entering the representation.
Physically, the $n_l$ fields of highest dimension are the auxiliary fields which transform as a time-derivative under any supersymmetry generator. The maximal value $l$ (corresponding to the maximal dimensionality
$d_l$) is defined to be the length of the representation (a root representation has length $l=2$).
Either $n_1, n_3,\ldots$ correspond to the bosonic fields (therefore $n_2, n_4, \ldots$ specify the fermionic fields)
or viceversa.  In both cases the equality $n_1+n_3+\ldots =n_2+n_4+\ldots = n$ is guaranteed.\par
The admissible fields contents were classified in \cite{krt}.
The complete list of the allowed fields contents was explicitly produced for all values $N\leq 10$.
Some corollaries follow from the \cite{krt} construction. $N=1,2,4,8$ are the only values of $N$ such that
all its irreducible representations have length $l\leq 3$. Conversely, starting from $N\geq 10$,
irreducible representations with length $l=5$ fields contents are allowed.\par
\par
A graphical presentation of the supersymmetry transformations was introduced in \cite{fg}.
It was later pointed out in \cite{dfghil} and \cite{dfghil2} (some specific $N=5,6$ examples were given) that the fields contents alone do not
necessarily uniquely specify the finite linear irreducible representations of (\ref{nsusyalgebra}).
This result was based on a notion of class of equivalence of the irreducible representations of
the $1D$ $N$-Extended Superalgebra motivated by the set of moves acting on its graphical presentations (see also the comments in \cite{top2}).\par
In a previous paper \cite{kt} we classified, up to $N\leq 8$, all inequivalent irreducible representations admitting the same fields content but differing in connectivity of their associated graphs. The notion of connectivity, describing
how vertices and edges are linked together, was made precise by the introduction of the so-called $\psi_g$ symbol, whose definition will be recalled in the next Section. A particular consequence of the \cite{kt} results is that, up to $N\leq 8$, inequivalent connectivities are only encountered for $N=5,6$.\par
In \cite{kt} no attempt was made to explain the motivation of the found results. In this paper we address this issue.
The $N=5$ irreducible supermultiplets contain twice as many fields than the $N=4$ irreducible supermultiplets
(\ref{irrepdim}). It is therefore natural to decompose a given $N=5$ supermultiplet into two $N=4$ supermultiplets whose respective fields are linked together by an extra, fifth, supersymmetry transformation.
The decomposition of an $N=5$ supermultiplet of given fields content can be performed
in different ways, according to the fields contents of the two composing $N=4$ supermultiplets.
Since the overall fields content of the $N=5$ supermultiplet is fixed, we find equalities (see (\ref{N4dec})) such as
$(n, 8, 8-n)= (k,4,4-k)+(n-k,4,4-n+k)$, where each inequivalent value of $k$ produces a different $N=4$ decomposition.
To each such decomposition, an inequivalent $\psi_g$ connectivity symbol for the overall $N=5$ supermultiplet is associated.
This result is not trivial. Without the analysis of the connectivity we could have easily thought that the different decompositions into $N=4$ supermultiplets (which can also be described, in the superfield language, as different decompositions in terms of $N=4$ superfields) would produce equivalent results. This is not the case. The analysis of the connectivity proves their inequivalence. Conversely, the whole list of $N=5$ inequivalent connectivities is completely specified by the inequivalent $N=4$ decompositions.  \par
This result opens the way to extend the analysis of the inequivalent connectivities to the $N=9$ irreducible supermultiplets. The reason is simple. They are constructed in terms of two irreducible $N=8$ supermultiplets
and an extra supersymmetry transformation connecting them. The complete list of $N=9$ inequivalent connectivities is reported in Section {\bf 4}. We further analyze the inequivalent connectivities of some $N=10$ irreducible supermultiplets which can be obtained from two irreducible $N=9$ supermultiplets linked together by an extra supersymmetry.\par
The results of this paper are further complemented by the presentation of the ``oxidation diagrams" which provide the following information. The irreducible $N=5,6,7,8$ supermultiplets admit the same number of fields. An oxidation diagram describes whether an $N=5$ irreducible representation can be lifted (``oxidized") to an $N=6$ (and to which $N=6$, in the case of fields contents with inequivalent connectivities), an $N=7$ or an $N=8$ irreducible representation.
All length-$2$ and length-$3$ $N=5$ irreducible representation can be oxidized to $N=8$ irreducible representations.
On the other hand, certain $N=5$ irreducible supermultiplets cannot be oxidized to all the inequivalent $N=6$
supermultiplets with same fields content, as shown in figures $3,4,5$.

\par
The structure of the paper is as follows. In the next Section, to make the paper self-consistent, the graphical presentation of the
supersymmetry transformations and the notion of connectivity is briefly recalled.
In Section {\bf 3} it is shown that the inequivalent irreducible $N=5$ supermultiplets with same fields content are
due to their different decompositions into two $N=4$ irreducible supermultiplets. This construction is extended in Section {\bf 4} to prove that the inequivalent $N=9$ irreducible supermultiplets with same fields content are obtained by their
inequivalent decompositions w.r.t. the $N=8$ superfields. In Section {\bf 5} inequivalent $N=10$ supermultiplets are constructed. In Section {\bf 6} the construction of oxidation diagrams, expressing the lifting of the irreducible $N=5$ supermultiplets to irreducible supermultiplets of the $N=6,7,8$ extended supersymmetries, is discussed. In the Conclusions some comments are made on various applications of these results. The paper is integrated by $5$ figures, two of them graphically depicting an example of decomposition of the supersymmetry transformations, while the three remaining ones present oxidation diagrams.

\section{Supersymmetric graphs and their connectivity}
Some technical issues which will be used for later computations (e.g. the dressing transformations which progressively lengthen the irreducible supermultiplets, the constraints that they have to satisfy, etc.)  have been discussed at length in \cite{krt} and \cite{kt}. To save space, they will not be repeated here.\par
In this Section we describe, largely based on \cite{fg}, the graphical interpretation of the irreducible supersymmetry transformations and discuss, based on \cite{kt}, their connectivity properties.
\par
An association can be made between $N$-colored oriented graphs and the linear supersymmetry
transformations. The identification goes as follows. The fields (bosonic and fermionic) entering a representation
are expressed as vertices. They can be accommodated into an $X-Y$ plane. The $Y$ coordinate  can be chosen to
correspond to the mass-dimension $d$ of the fields. Conventionally, the lowest dimensional fields can be
associated to vertices lying on the $X$ axis. The higher dimensional fields have positive, integer or half-integer values of $Y$.
A colored edge links two vertices which are connected by a supersymmetry transformation. Each one of the $N$ $Q_i$ supersymmetry generators is associated to a given color. The edges are oriented. The orientation reflects the sign
(positive or negative) of the corresponding supersymmetry transformation connecting the two vertices. Instead of using
arrows, alternatively, solid or dashed lines can be associated, respectively, to positive or negative signs.
No colored line is drawn for supersymmetry transformations connecting a field with the time-derivative of a lower
dimensional field. This is in particular true for the auxiliary fields (the fields of highest dimension in the representation) which are necessarily mapped, under supersymmetry transformations, in the time-derivative of lower-dimensional fields.
\par
Each irreducible supersymmetry transformation can be presented (the identification is not unique) through an oriented
$N$-colored graph with $2n$ vertices (see (\ref{irrepdim})). The graph is such that precisely $N$ edges, one for each
color, are linked to any given vertex which represents either a $0$-mass dimension or a $\frac{1}{2}$-mass dimension field. \par
Despite the fact that the presentation of the graph is not unique, certain of its features only depend on the class
of the supersymmetry transformations. We introduce now, following \cite{kt}, the invariant characterization. An unoriented ``color-blind" graph can be associated to the initial graph by disregarding the orientation of the edges
and their colors (all edges are painted in black). For simplicity, we discuss here the invariant characterization
of the graphs associated to the length $l=3$ irreducible representation that will be discussed in the following (the generalization of the invariant characterization to graphs of arbitrary length is straightforward, see \cite{kt}). They admit fields content $(k,n,n-k)$. The corresponding fields are denotes as $x_p$ (for $0$-mass dimension),
$\psi_q$ (for $\frac{1}{2}$ mass-dimension) and $g_r$ (the $1$ mass-dimension auxiliary fields), where $p=,1,\ldots, k$, $q=1,\ldots , n$ and $r=1,\ldots, n-k$.\par
The connectivity of the associated length $l=3$ color-blind graph can be expressed through the connectivity symbol
$\psi_g$, expressed as
\begin{eqnarray}
\psi_g &=& ({m_1})_{s_1} +({m_2})_{s_2}+\ldots +({m_Z})_{s_Z}.
\end{eqnarray}
The $\psi_g$ symbol encodes the information on the partition of the $n$  $\frac{1}{2}$-mass dimension fields (vertices)
into the sets of $m_z$ vertices ($z=1,\ldots, Z$) with $s_z$ edges connecting them to the $n-k$ $1$-mass dimension auxiliary fields.
We have
\begin{eqnarray}
m_1+m_2+\ldots +m_Z &=& n,
\end{eqnarray}
while $
s_z\neq s_{z'}$ for $  z\neq z'$.\par
The connectivity symbol is an invariant characterization of the class of the irreducible supersymmetry transformations.\par
The connectivity symbol $\psi_g$ can be used to induce a map ${\widetilde{\psi_g}}$ from the set of graphs $Gr$ into the set of integers ${\bf Z}$
(${\widetilde \psi_g}: Gr\rightarrow {\bf Z}$) s.t. $W\in {\bf Z}$ is given by
\begin{eqnarray}
W&=&\prod_{z=1}^{Z}(p_{2z-1}^{m_z})(p_{2z}^{s_z}),
\end{eqnarray}
where the $p_w$'s, $w=1,2,3, \ldots$, denote the ordered set of prime integers ($2,3,5,\ldots$). With the above definition
two inequivalent connectivities induce two distinct integers $W,W'$ ($W'\neq W$).

\section{$N=4$ decompositions and connectivities of the $N=5$ supermultiplets}

The $N=5$ irreducible supermultiplets contain a total number of $8$ bosonic and $8$ fermionic fields.
The $N=5$ supermultiplets can be decomposed into two sets of $N=4$ irreducible supermultiplets (superfields),
whose vertices (component fields) are linked together by the $5^{th}$ supersymmetry. The $N=4$ superfields
contain $4$ bosonic and $4$ fermionic fields associated to different mass-dimensions.
\par
The length-$2$ and length-$4$ $N=5$ supermultiplets admit a unique decomposition into $N=4$ supermultiplets.
The situation is different for the length-$3$ $N=5$ supermultiplets whose fields content is given by
$(n, 8, 8-n)$, for $n=1,2,\ldots, 7$. They admit the following decompositions in terms of
$(k,4,4-k)$ and $(n-k,4,4-n+k)$ $N=4$ supermultiplets:
\begin{eqnarray}\label{N4dec}
(n, 8, 8-n)&=& (k,4,4-k)+(n-k,4,4-n+k).
\end{eqnarray}

It is convenient to express $n$ as
\begin{eqnarray}
n &=& 4+\epsilon m,
\end{eqnarray}
where $\epsilon=\pm 1$, while $m=0,1,2,3$. \par
The inequivalent values of $k$ are given by the integers
\begin{eqnarray}
k&=& \frac{1}{2}(1+\epsilon)m, \frac{1}{2}(1+\epsilon)m +1,\ldots, \frac{1}{2}(1+\epsilon) m +[\frac{4-m}{2}],
\end{eqnarray}
where the square brackets refers to the integral part.\par
The $\psi_g$ connectivity symbol can be easily computed for each such decomposition. We obtain, in terms of $n$ and $k$,
\begin{eqnarray}
\psi_g &=&{(4-k)}_{5+k-n}+(k)_{4+k-n}+(4+k-n)_{5-k}+(n-k)_{4-k}.
\end{eqnarray}
For any given $n$, the $\psi_g$ connectivity symbol differs for inequivalent values of $k$.
This implies, as a corollary, that the decomposition into $N=4$ supermultiplets specified by different, inequivalent
values of $k$ produces inequivalent $N=5$ irreducible supermultiplets (no matter which supersymmetry generator is picked up as the ``fifth").\par
We define as ``$\Delta$" the number of degeneracies, i.e. the number of inequivalent supermultiplets with the same fields content. $\Delta$ is computed to be
\begin{eqnarray}
\Delta &=& [\frac{4-m}{2}]+1,
\end{eqnarray}

The results for the inequivalent $N=5$ length-$3$ supermultiplets can be summarized in the following table
\begin{eqnarray}
\begin{tabular}{|l|l|l|c|}
\hline
fields cont. & $N=4$ decomp. & $\psi_g$ connectivities  & labels\\
\hline
$(1,8,7)$  & $(0,4,4)+(1,4,3)$ & $3_5+5_4$ &\\
\hline
$(2,8,6)$  & $(0,4,4)+(2,4,2)$ & $2_5+2_4+4_3$&$A$ \\
  & $(1,4,3)+(1,4,3)$ & $6_4+2_3$ &$B$\\
\hline
$(3,8,5)$  & $(0,4,4)+(3,8,5)$ & $1_5+3_4+4_2$ &$A$\\
  & $(1,4,3)+(2,4,2)$ & $2_4+5_3+1_2$ &$B$\\
\hline
$(4,8,4)$  & $(0,4,4)+(4,4,0)$ & $4_4+4_1$ &$A$\\
  & $(1,4,3)+(3,4,1)$ & $1_4+3_3+3_2+1_1$ &$B$\\
   & $(2,4,2)+(2,4,2)$ & $4_3+4_2$ &$C$\\
   \hline
   $(5,8,3)$  & $(1,4,3)+(4,4,0)$ & $4_3+3_1+1_0$ &$A$\\
  & $(2,4,2)+(3,4,1)$ & $1_3+5_2+2_1$&$B$ \\
\hline
$(6,8,2)$  & $(2,4,2)+(4,4,0)$ & $4_2+2_1+2_0$&$A$ \\
  & $(3,4,1)+(3,4,1)$ & $2_2+6_1$ &$B$\\
\hline
$(7,8,1)$  & $(3,4,1)+(4,4,0)$ & $5_1+3_0$& \\

\hline
\end{tabular}
\end{eqnarray}
The last column specifies the labels assigned, in terms of increasing values of $k$, to each inequivalent $N=5$ supermultiplet.

For the sake of clarity we show here two unoriented graphs (the figures $1$ and $2$ in the tables attached at the end of the paper) associated to the two inequivalent $N=5$ supermultiplets ($A$ and respectively $B$) with same fields content
$(2,8,6)$. Contrary to the graphs already presented in the literature (see \cite{dfghil2} and \cite{kt}), the figures $1$ and $2$ stress the $N=4$ decompositions of the overall $N=5$ graphs. Similar graphical presentations can be straightforwardly drawn for all other inequivalent $N=5$ supermultiplets.

\section{$N=8$ decompositions and connectivities of the  $N=9$ supermultiplets}

The treatment of the inequivalent $N=9$ irreducible supermultiplets is made in parallel with the $N=5$ case.
The reason is that the $N=9$ supermultiplets contain $16$ bosonic and $16$ fermionic fields which can be decomposed into two sets of $N=8$ irreducible supermultiplets linked together by a $9^{th}$ supersymmetry.
Just like the $N=5$ case, the $N=9$ length-$2$ and length-$4$ irreducible supermultiplets (given in \cite{krt})
admit a unique decomposition into $N=8$ supermultiplets. For length-$3$ supermultiplets
$(n,16,16-n)$, with $n=1,2,\ldots, 15$, we have the decompositions
\begin{eqnarray}
(n, 16, 16-n)&=& (k,8,8-k)+(n-k,8,8-n+k).\end{eqnarray}
We express $n$ as
\begin{eqnarray}
n &=& 8+\epsilon m,
\end{eqnarray}
where $\epsilon=\pm 1$, while $m=0,1,\ldots, 7$. \par
The inequivalent values of $k$ are given by the integers
\begin{eqnarray}
k&=& \frac{1}{2}(1+\epsilon)m, \frac{1}{2}(1+\epsilon)m +1,\ldots, \frac{1}{2}(1+\epsilon) m +[\frac{8-m}{2}].
\end{eqnarray}

The computation of the $\psi_g$ connectivity symbol gives us
\begin{eqnarray}
\psi_g &=&{(8-k)}_{9+k-n}+(k)_{8+k-n}+(8+k-n)_{9-k}+(n-k)_{8-k}.
\end{eqnarray}
At a fixed value of $n$ we obtain distinct $\psi_g$ connectivity symbols for inequivalent values of $k$.\par
The ``degeneracies number" $\Delta$ is now given by
\begin{eqnarray}
\Delta &=& [\frac{8-m}{2}]+1.
\end{eqnarray}
The results for the $N=9$ length-$3$ supermultiplets are summarized in the table
\begin{eqnarray}
\begin{tabular}{|l|l|l|c|}\hline
f.lds con. & $N=8$ decomp. & $\psi_g$ connectivities  & labels\\
\hline
$(1,16,15)$  & $(0,8,8)+(1,8,7)$ & $7_9+9_8$ &\\
\hline
$(2,16,14)$  & $(0,8,8)+(2,8,6)$ & $6_9+2_8+8_7$&$I$ \\
  & $(1,8,7)+(1,8,7)$ & $14_8+2_7$ &$II$\\
\hline
$(3,16,13)$  & $(0,8,8)+(3,8,5)$ & $5_9+3_8+8_6$ &$I$\\
  & $(1,8,7)+(2,8,6)$ & $6_8+9_7+1_6$ &$II$\\
\hline
$(4,16,12)$  & $(0,8,8)+(4,8,4)$ & $4_9+4_8+8_5$ &$I$\\
  & $(1,8,7)+(3,8,5)$ & $5_8+3_7+7_6+1_5$ &$II$\\
   & $(2,8,6)+(2,8,6)$ & $12_7+4_6$ &$III$\\
   \hline
   $(5,16,11)$  & $(0,8,8)+(5,8,3)$ & $3_9+5_8+8_4$ &$I$\\
  & $(1,8,7)+(4,8,4)$ & $4_8+4_7+7_5+1_4$&$II$ \\
   & $(2,8,6)+(3,8,5)$ & $5_7+9_6+2_5$ &$III$\\
\hline
$(6,16,10)$  & $(0,8,8)+(6,8,2)$ & $2_9+6_8+8_3$&$I$ \\
  & $(1,8,7)+(5,8,3)$ & $3_8+5_7+7_4+1_3$ &$II$\\
   & $(2,8,6)+(4,8,4)$ & $4_7+4_6+6_5+2_4$ &$III$\\
     & $(3,8,5)+(3,8,5)$ & $10_6+6_5$&$IV$ \\
\hline
$(7,16,9)$  & $(0,8,8)+(7,8,1)$ & $1_9+7_8+8_2$&$I$ \\
  & $(1,8,7)+(6,8,2)$ & $2_8+6_7+7_3+1_2$ &$II$\\
   & $(2,8,6)+(5,8,3)$ & $3_7+5_6+6_4+2_3$ &$III$\\
     & $(3,8,5)+(4,8,4)$ & $4_6+9_5+3_4$&$IV$ \\
\hline
$(8,16,8)$  & $(0,8,8)+(8,8)$ & $8_8+8_1$&$I$ \\
 & $(1,8,7)+(7,8,1)$ & $1_8+7_7+7_2+1_1$ &$II$\\
  & $(2,8,6)+(6,8,2)$ & $2_7+6_6+6_3+2_2$ &$III$\\
   & $(3,8,5)+(5,8,3)$ & $5_5+5_4+3_6+3_3$ &$IV$\\
     & $(4,8,4)+(4,8,4)$ & $8_5+4_4$&$V$ \\
\hline
$(9,16,7)$ & $(1,8,7)+(8,8)$ & $8_7+7_1+1_0$ &$I$\\
 & $(2,8,6)+(7,8,1)$ & $1_7+7_6+6_2+2_1$ &$II$\\
  & $(3,8,5)+(6,8,2)$ & $2_6+6_5+5_3+3_2$ &$III$\\
   & $(4,8,4)+(5,8,3)$ & $3_5+5_4+4_3$ &$IV$\\
\hline
$(10,16,6)$ & $(2,8,4)+(8,8)$ & $6_6+6_1+2_0$ &$I$\\
 & $(3,8,5)+(7,8,1)$ & $1_6+7_5+5_2+3_1$ &$II$\\
  & $(4,8,4)+(6,8,2)$ & $2_5+6_4+4_3+4_2$ &$III$\\
   & $(5,8,3)+(5,8,3)$ & $6_4+10_3$ &$IV$\\
\hline
$(11,16,5)$ & $(3,8,5)+(8,8)$ & $8_5+5_1+3_0$ &$I$\\
 & $(4,8,4)+(7,8,1)$ & $ 1_5+7_4+4_2+4_1$ &$II$\\
  & $(5,8,3)+(6,8,2)$ & $5_4+9_3+5_2$&$III$ \\
\hline
$(12,16,4)$ & $(4,8,4)+(8,8)$ & $8_4+4_1+4_0$ &$I$\\
 & $(5,8,3)+(7,8,1)$ & $1_4+7_3+3_2+5_1$&$II$ \\
  & $(6,8,2)+(6,8,2)$ & $4_3+12_2$&$III$ \\
\hline
$(13,16,3)$ & $(5,8,3)+(8,8)$ & $8_3+3_1+5_0$&$I$ \\
 & $(6,8,2)+(7,8,1)$ & $1_3+9_2+6_1$ &$II$\\
\hline
$(14,16,2)$ & $(6,8,2)+(8,8)$ & $8_2+2_1+6_0$ &$I$\\
 & $(7,8,1)+(7,8,1)$ & $2_2+14_1$ &$II$\\
\hline
$(15,16,1)$ & $(7,8,1)+(8,8)$ & $9_1+7_0$ &\\
\hline
\end{tabular}\label{N9conn}
\end{eqnarray}

We use roman numerals (in the last column) to label inequivalent $N=9$ supermultiplets with same fields content.\par

Like their $N=5$ counterparts, inequivalent $N=9$ supermultiplets can be straightforwardly presented in terms of graphs
stressing their decompositions in terms of $N=8$ superfields.

\section{On decompositions and connectivities of the $N=10$ supermultiplets}

The irreducible $N=10$ supermultiplets contain $32$ bosonic and $32$ fermionic fields. They can be decomposed
into two $N=9$ irreducible supermultiplets or four $N=8$ irreducible supermultiplets.
We describe here their decomposition into $N=9$ supermultiplets with an extra supersymmetry connecting the fields belonging to the two different $N=9$ supermultiplets.\par
For simplicity we limit our discussion here to the length-$3$ $N=10$ irreducible supermultiplets with fields content
$(k,32,32-k)$ and $k$ given by $k=1,2, 3$ (the general case is conducted along the same lines).\par
For $k=1$ the decomposition into $N=9$ supermultiplets is given by
\begin{eqnarray}
 (1,32,31)&=&(1,16,15)+(0,16,16).
 \end{eqnarray}
For $k=2$ we have three types of decompositions, namely
\par
{\em i}) $(2,16,14)_I+(0,16,16)$,\par
{\em ii}) $(2,16,14)_{II}+ (0,16,16)$,\par
{\em iii}) $(1,16,15)+(1,16,15)$.
\par
The $\psi_g$ connectivity symbol can be easily computed and the following results are encountered:
\begin{eqnarray}
(1,32,31):  &\psi_g\equiv& 20_{10}+10_9,
\nonumber\\
 (2,32,31)_{case ~i}:  &\psi_g\equiv& 20_{10}+4_9+8_8,\nonumber \\
 (2,32,31)_{case ~ii}:  &\psi_g\equiv& 14_{10}+16_9+2_8.
 \end{eqnarray}
 In the third case {\em iii}) we are faced for the first time with a new feature,
 the fact that the decomposition into
$N=9$ supermultiplets doesf not determine, alone, the connectivity of the $N=10$ supermultiplet.
Indeed, the two $N=9$ $(1,16,15)$ supermultiplets admit $N=9$ $\psi_g$ connectivity given by
$9_8+7_9$. The extra, $10^{th}$ supersymmetry, can be implemented in two inequivalent ways. In the case $\alpha$
the unique $0$-mass
dimension field of the first $N=9$ supermultiplet is linked with one of the $7$ fermionic fields of connectivity $9$  of the second supermultiplet (as a consequence, the unique $0$-mass
dimension field of the second $N=9$ supermultiplet is linked with one of the $7$ fermionic fields of connectivity $9$
of the first supermultiplet). In the case $\beta$
the unique $0$-mass
dimension field of the first $N=9$ supermultiplet is linked with one of the $9$ fermionic fields of connectivity $8$  of the second supermultiplet (as a consequence, the unique $0$-mass
dimension field of the second $N=9$ supermultiplet is linked with one of the $9$ fermionic fields of connectivity $8$
of the first supermultiplet).
\par
The $\beta$ case produces the same $\psi_g$ connectivity symbol as the one given by $(2,32,31)_{case ~ii} $.
On the other hand,
the $\alpha$ case produces a new, inequivalent, $\psi_g$ connectivity symbol given by
\begin{eqnarray}
(2,32,31)_{case ~iii,\alpha}: &\psi_g\equiv& 12_{10}+20_9.
\end{eqnarray}
The list of the admissible $N=10$  $\psi_g$ connectivity symbols for the $(3,32,29)$ fields content is given by
\begin{eqnarray}
(3,32,29)_{I}: &\psi_g\equiv& 18_{10}+6_9+8_7, \nonumber\\
(3,32,29)_{II}: &\psi_g\equiv& 13_{10}+9_9+9_8+1_7, \nonumber\\
(3,32,29)_{III}: &\psi_g\equiv& 12_{10}+10_9+10_8, \nonumber\\
(3,32,29)_{IV}: &\psi_g\equiv& 10_{10}+14_9+8_8, \nonumber\\
(3,32,29)_{V}: &\psi_g\equiv& 7_{10}+21_9+3_8+1_7, \nonumber\\
(3,32,29)_{VI}: &\psi_g\equiv& 6_{10}+22_9+4_8.
\end{eqnarray}

\section{Oxidation properties of the $N=5,6,7,8$ supermultiplets}

The analysis of the algebraic understanding of the connectivities would not be complete without the investigation of the liftings of the $N=5$ supermultiplets to $N=6,7,8$ supermultiplets. This situation differs from the cases treated so far
since all irreducible supermultiplets of the $N=5,6,7,8$ extended supersymmetries admit the same number (equal to $8$)
of bosonic (and fermionic) fields. The lifting $N'\rightarrow N''$ of an irreducible representation of the $N'$-extended supersymmetry to an $N''>N'$-extended supersymmetry will be referred to as ``oxidation". Originally, in the superstring/$M$-theory literature, the term oxidation was a pun applied to the lifting of a lower-dimensional theory to a higher-dimensional one.
The application of this term in our context is justified since a supersymmetric theory in higher-dimension produces, when dimensionally reduced, an extended supersymmetric quantum mechanical model in one time dimension. For instance,
the $N=2$ $D=4$
superfields entering the $N=2$ Super-Yang-Mills theory project to irreducible supermultiplets of the $N=8$ Supersymmetric Quantum Mechanics (see, e.g., the discussion in \cite{top}).\par
The complete list of the $N=5,6,7,8$ irreducible supermultiplets have been produced in \cite{kt}. Inequivalent connectivities are only found for certain length-$3$ supermultiplets of the $N=5,6$ extended supersymmetries.
The inequivalent $N=5$ irreducible supermultiplets with same fields content have been rederived in Section {\bf 3}.
The inequivalent $N=6$ irreducible supermultiplets, specified by their $\psi_g$ connectivity symbol and label, are listed below:
\begin{eqnarray}
\begin{tabular}{|l|l|c|}
\hline
f.lds con. & $\psi_g$ connectivities  & labels\\
\hline
$(2,8,6)$  & $2_{6}+6_4$ & $A$\\
$(2,8,6)$  & $4_5+4_4$ & $B$\\
\hline
$(3,8,5)$  & $2_5+2_4+4_3$ & $A$\\
$(3,8,5)$  & $6_4+2_3$ & $B$\\
\hline
$(4,8,4)$  & $4_4+4_2$ & $A$\\
$(4,8,4)$  & $8_3$ & $B$\\
$(4,8,4)$  & $2_4+4_3+2_2$ & $C$\\
\hline
$(5,8,3)$  & $4_3+2_2+2_1$ & $A$\\
$(5,8,3)$  & $2_3+6_2$ & $B$\\
\hline
$(6,8,2)$  & $6_2+2_0$ & $A$\\
$(6,8,2)$  & $4_2+4_1$ & $B$\\
\hline
\end{tabular}
\end{eqnarray}
When the fields content uniquely specifies the irreducible $N=5$ supermultiplet, the oxidation is given
by the following diagrams (the maximal value of the extended $N$, the ``oxidized supersymmetry", is underlined):
\begin{eqnarray}
(1,5,7,3),~(3,7,5,1)  &:& {\underline{N=5}},\nonumber\\
(1,6,7,2),~(2,6,6,2), ~(2,7,6,1) &:& N=5\rightarrow {\underline{N=6}},\nonumber\\
(1,7,7,1) &:& N=5\rightarrow N=6\rightarrow {\underline{N=7}},\nonumber\\
(8,8),~ (1,8,7), ~(7,8,1) &:& N=5\rightarrow N=6\rightarrow N=7\rightarrow  {\underline{N=8}}.
\end{eqnarray}
For what concerns the irreducible supermultiplets with inequivalent connectivities, the corresponding oxidation diagrams are given in figures $3,4,5$ in the tables.
In each diagram a dot specifies the  $N=5,6,7,8$ irreducible supermultiplets (together with their label)  of a given fields content. A line is drawn between dots (supermultiplets) admitting an $N'\rightarrow N'+1$ lifting.
All supermultiplets can be oxidized to the maximal $N=8$ supersymmetry. On the other hand, the absence of a line connecting, e.g.,
$N=5$ $(2,8,6)_B$ and $N=6$ $(2,8,6)_A$ in figure $4$, means that one cannot directly lift $N=5$ $(2,8,6)_B$ to $N=6$ $(2,8,6)_A$.
\par
Notice that the oxidation diagrams respect the duality property of the irreducible supermultiplets
$(n_1,n_2,\ldots , n_l)\leftrightarrow (n_l, \ldots,
n_2,n_1)$ discussed in \cite{krt} and \cite{kt}.

\section{Conclusions}

In the Conclusions we discuss various setups where the results of this paper can find applications.
We mention at first the existence of constraints for higher-dimensional supersymmetric theories. These constraints can be recovered
from the
dimensional
reduction to one-dimensional supersymmetric quantum mechanical systems (see \cite{top} for a broader review).
In this context, various identities in \cite{bbbm} and the works there cited find a natural explanation in terms of the representation theory of the $1D$ $N$-extended supersymmetry algebra (\ref{nsusyalgebra}). The $10$ dimensional Super-Yang-Mills theory admits $9$ off-shell supersymmetries (the $7$ extra supersymmetries which add to the total number of $N=16$ supersymmetries close on-shell). The supermultiplet carrying the $N=9$ supersymmetry representation is given \cite{ber} by $9$ physical bosons, $16$ fermions and $7$ additional auxiliary bosonic degrees of freedom. It corresponds to the
length-$3$ $(9,16,7)$ field content of $N=9$. We proved in this work the existence of four inequivalent $N=9$ irreducible
representations which correspond to the $(9,16,7)$ field content. Since the seven auxiliary fields in \cite{ber}
are associated to the imaginary octonions and are all on equal footing, it turns out that the irreducible supermultiplet
corresponding to the $10$-dimensional SYM theory is the $(9,16,7)_I$ of formula (\ref{N9conn}) (namely, the one obtained
from the $(1,8,7)+(0,8,8)$ decomposition in $N=8$ supermultiplets). \par
The physical content of the $11$-dimensional supergravity is given \cite{cjs} by a set of fields entering different representations
of the $so(9)$ transverse algebra. We have $44$ fields associated to the traceless symmetric graviton, $128$ fermionic
fields for the gravitino's degrees of freedom, $84$ fields for the antisymmetric $3$-form. They give rise to a length-$3$ irreducible multiplet of field content $(44,128,84)$. It carries (see (\ref{irrepdim})) an off-shell representation of  $N=16$ supersymmetries (the remaining $16$ supersymmetries which add up to the total number of $N=32$ supersymmetries of the maximal supergravity, close on-shell). In this work we have shown that,
starting from $N\geq 9$, inequivalent length-$3$ irreducible representations with the same field content can be
found. The correct identification of the $(44,128,84)$ irreducible representation of $N=16$ which corresponds to the
$11$-dimensional supergravity multiplet has to be determined.\par
It is worth to spend some words on the connection of the irreducible supermultiplets here discussed and the superfield formalism. In \cite{abc} the superfield descriptions for all the irreducible supermultiplets of the $N=8$ extended supersymmetry were produced (the $N=4$ superfield descriptions can be found in the references given in \cite{abc}).
For low ($N\leq 8$) values of $N$ the analysis of the irreducible representations is essential in order to determine the complete set of superfields and the constraints that their component fields have to satisfy. In any case, both the superfields formalism (once introduced) and the informations contained in the irreducible supermultiplets are equivalent, even in the construction of supersymmetric invariant actions.
This was proven in \cite{krt}, where a new $N=8$ off-shell invariant action, later discussed also in \cite{di}, was found by taking into account the properties of the $N=8$ irreducible representations, without explicitly using the superfield formalism.
On the other hand, for large values of $N$, the superfield formalism becomes impractical, while the irreducible representations of the $N$-extended supersymmetries are available and algorithmically constructed with the method described in \cite{krt}. Let us discuss a relevant example.
An off-shell formulation of the $N=32$ supersymmetric quantum mechanics requires (see (\ref{irrepdim})) $32768$ bosonic,
as well as an equal number of fermionic, fields. A corollary of the (\ref{irrepdim}) formula is the existence of a necessary condition on the minimal number ($
32768+32768$) of fields which would be required to produce an off-shell formulation of the $11$-dimensional maximal supergravity or the $M$-theory. A superfield description with $32$ Grassmann parameters is highly reducible, containing
$2^{31}$ bosonic and $2^{31}$ fermionic degrees of freedom. To extract from them an irreducible representation, more than $4\times 10^9$ constraints have to be imposed. These constraints have to be imposed for each inequivalent
$N=32$ irreducible supermultiplet (we recall that there are $32767$ inequivalent fields contents of length-$3$).  \par
The supersymmetry transformations investigated in this paper are linear. In the literature, non-linear realizations of the $N$ extended supersymmetry algebra (\ref{nsusyalgebra}) are also discussed. As recalled in \cite{bkls}, the one-dimensional supersymmetric off-shell invariant actions
constructed with linear representations of the $N$ extended supersymmetry (such as the ones given by supermultiplets with $(k,n,n-k$) field content), correspond to sigma-models whose $k$-dimensional target manifold possesses a conformally flat metric. Supersymmetric, non-conformally flat sigma-models are obtained from the non-linear realizations of the (\ref{nsusyalgebra}) supersymmetry algebra. At present, the nature of the non-linear supersymmetry transformations is not
understood, since they have been produced using a large variety of methods. Recently, an interesting attempt at unifying these procedures has been made in \cite{bkls}. It amounts to produce non-linear realizations of the $N=4$ extended supersymmetry by suitably constraining the fields belonging to two $N=4$ extended irreducible linear supermultiplets.
Some similarities of this construction with the decomposition of an irreducible $N=5$ supermultiplet in terms of two
$N=4$ irreducible supermultiplets should be noticed.\par

\renewcommand{\theequation}{A.\arabic{equation}}
\setcounter{equation}{0}

{}~
\\{}~
\par {\large{\bf Acknowledgments}}{} ~\\{}~\par
Z. K. acknowledges FAPEMIG for financial support.

\newpage

{\Large{\bf Tables:}}
~\\
~\\

\begin{figure}[htbp]
%\centering{\scalebox{0.7}%{\includegraphics{Fig10}}
\epsfig{file=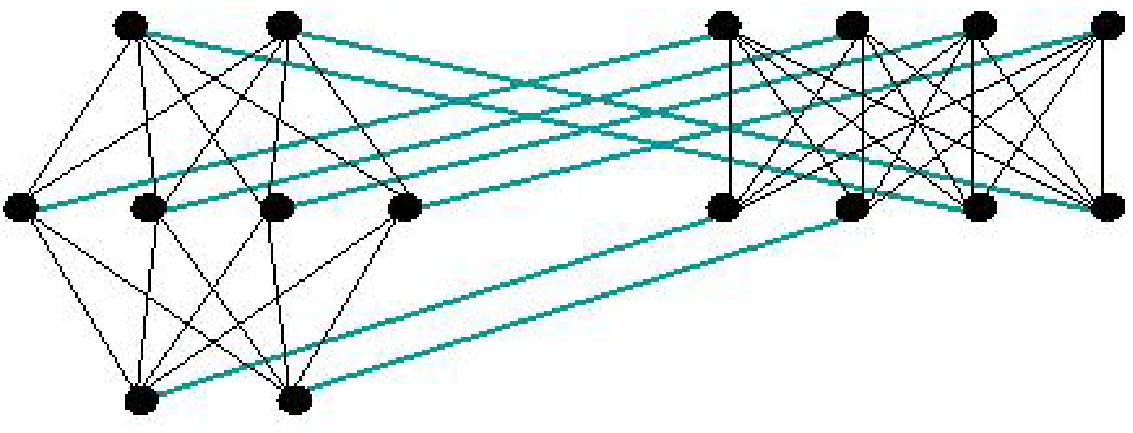}
%\centerline{\includegraphics{figura1.eps}}
%\vspace{-5.2 cm}
%\vspace{3 cm}
\caption{presentation of the $N=5$ $(2,8,6)_A$ supermultiplet as unoriented graph. It shows its decomposition into two
$N=4$ irreducible multiplets, $(2,4,2)$ and $(0,4,4)$. The $N=4$ supersymmetry transformations are drawn with black edges. The $5^{th}$ supersymmetry, connecting vertices belonging to the left and right $N=4$ supermultiplets, is given by the blue edges.}
\end{figure}
%\newpage
\begin{figure}[htbp]
%\centering{\scalebox{0.7}%{\includegraphics{Fig10}}
\epsfig{file=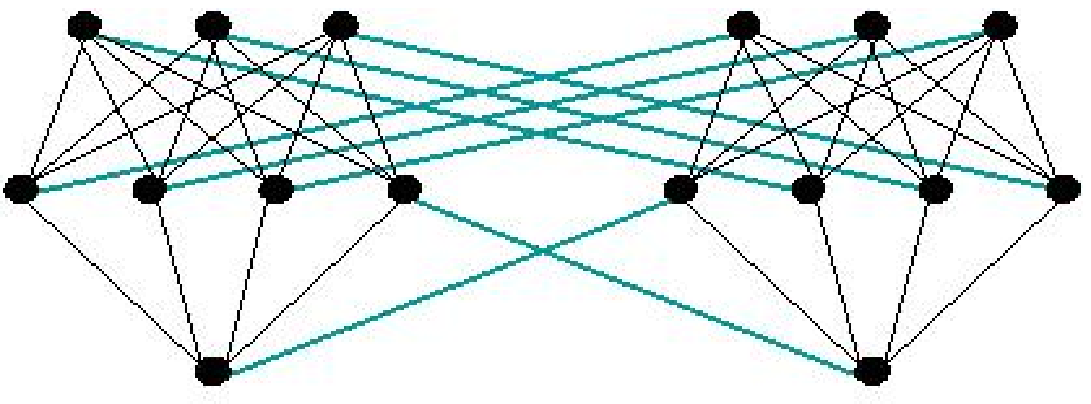}
%\centerline{\includegraphics{figura1.eps}}
%\vspace{-2.2 cm}
%\vspace{3 cm}
\caption{presentation of the $N=5$ $(2,8,6)_B$ supermultiplet as unoriented graph. It shows its decomposition into two
$N=4$ irreducible multiplets of $(1,4,3)$ type. The $N=4$ supersymmetry transformations are drawn with black edges. The $5^{th}$ supersymmetry, connecting vertices belonging to the left and right $N=4$ supermultiplets, is given by the blue edges.}
\end{figure}
%\newpage
\begin{figure}[htbp]
%\centering{\scalebox{0.7}{\includegraphics{Fig10}}
\epsfig{file=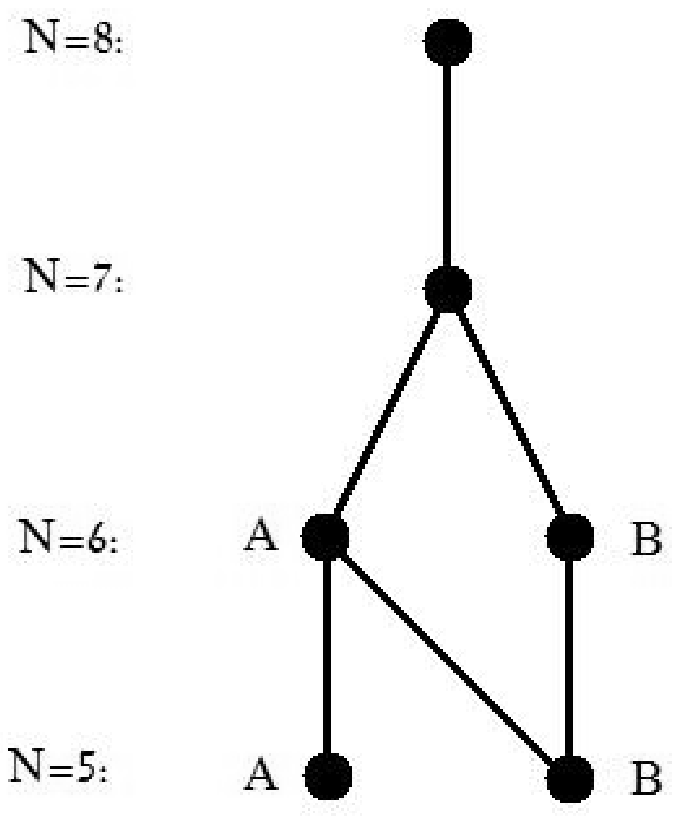}
%\centerline{\includegraphics{figura1.eps}}
%\vspace{-2.2 cm}
%\vspace{3 cm}
\caption{${\underline{N=5}}\Rightarrow {\underline{N=6}}\Rightarrow {\underline{N=7}} \Rightarrow {\underline{N=8}}$ oxidation diagram of the dually related irreducible supermultiplets
with $(2,8,6)$ and $(6,8,2)$ fields content. It can be read in reverse order as a reduction diagram.
The identification of the $N=5,6$ supermultiplets with $A,B$ labels and the connectivity of the irreps graphs (related to the decomposition into $N=4$ supermultiplets) is explained in the main text.}
%\vspace{1 cm}
%\end{figure}
\end{figure}
%\newpage
\begin{figure}
\epsfig{file=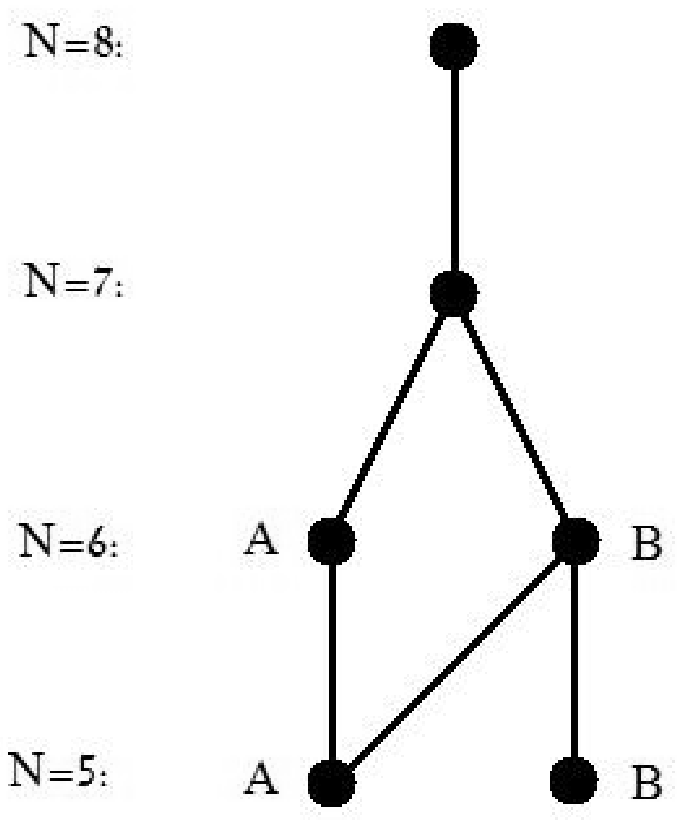}
%\centerline{\includegraphics{figura1.eps}}
%\vspace{-2.2 cm}
%\vspace{3 cm}
\caption{${\underline{N=5}}\Rightarrow {\underline{N=6}}\Rightarrow {\underline{N=7}} \Rightarrow {\underline{N=8}}$ oxidation diagram of the dually related irreducible supermultiplets
with $(3,8,5)$ and $(5,8,3)$ fields content. It can be read in reverse order as a reduction diagram.
The identification of the $N=5,6$ supermultiplets with $A,B$ labels and the connectivity of the irreps graphs (related to the decomposition into $N=4$ supermultiplets) is explained in the main text.}
\end{figure}
\newpage
\begin{figure}
%\end{figure}
%\begin{figure}[htbp]
%\centering{\scalebox{0.7}%{\includegraphics{Fig10}}
\epsfig{file=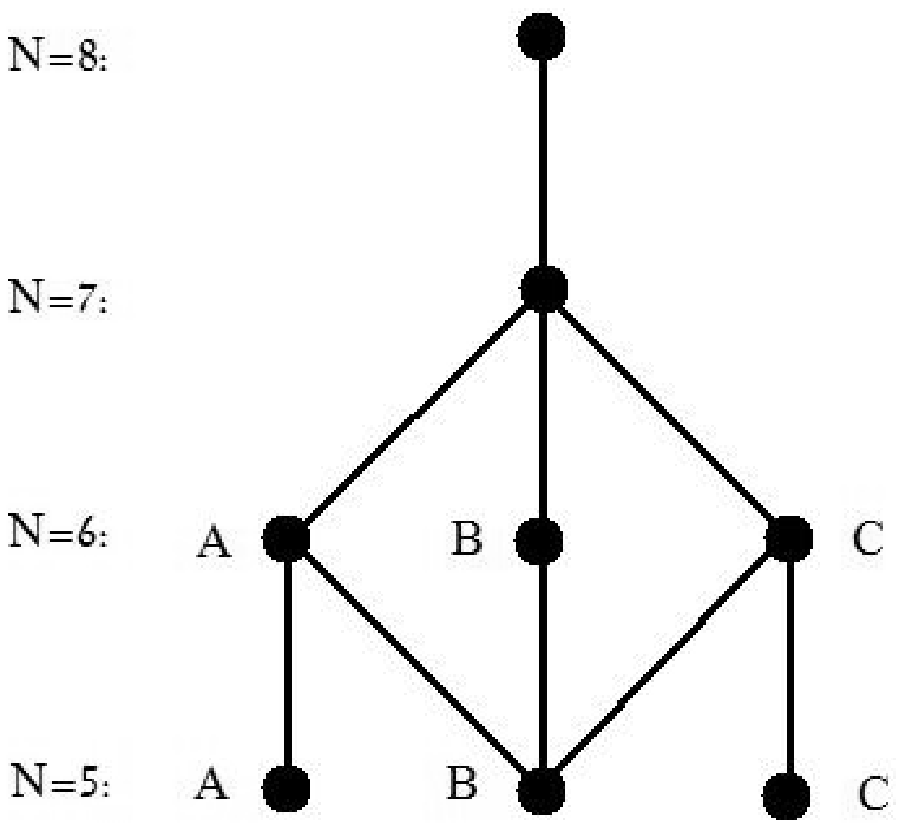}
%\centerline{\includegraphics{figura1.eps}}
%\vspace{-2.2 cm}
%\vspace{3 cm}
\caption{${\underline{N=5}}\Rightarrow {\underline{N=6}}\Rightarrow {\underline{N=7}} \Rightarrow {\underline{N=8}}$ oxidation diagram of the selfdual irreducible supermultiplets
with $(4,8,4)$ fields content. It can be read in reverse order as a reduction diagram.
The identification of the $N=5,6$ supermultiplets with $A,B,C$ labels and the connectivity of the irreps graphs (related to the decomposition into $N=4$ supermultiplets) is explained in the main text.}
\end{figure}

\end{document}